\documentclass[referee]{raa}            

\usepackage{graphicx,times}             

\begin{document}

   \title{An explanation for the peculiar periphery of supernova remnant G309.2-0.6
}

   \volnopage{Vol.0 (20xx) No.0, 000--000}      
   \setcounter{page}{1}          

   \author{Huan Yu
      \inst{1,2}
    \and Jun Fang
      \inst{2}   }

   \institute{Department of Physical Science and Technology, Kunming University, Kunming 650214, China; {\it yuhuan.0723@163.com}\\
        \and
             Department of Astronomy, Key Laboratory of Astroparticle Physics of Yunnan Province, Yunnan University, Kunming 650091, China; {\it fangjun@ynu.edu.cn}\\
   }

   \date{Received~~2018 month day; accepted~~2018~~month day}

\abstract{ Supernova remnant (SNR) G309.2-0.6 has a peculiar radio morphology with two bright ears to the southwest and northeast, although the main shell outside the ears is roughly circular. Based on an earlier proposal that the supernova ejecta has a jet component with extra energy,  the dynamical evolution of the remnant is solved using 3D hydrodynamical (HD) simulation to investigate the formation of the periphery of the remnant. Assuming the ejecta with a kinetic energy of $10^{51}$\,erg and a mass of $3~\mathrm{M}_{\odot}$ evolved in a uniform ambient medium for a time of $\sim 4000$\,yr and the jet component has cylindrical symmetry with a half open angle of $10^\circ$, the result indicates that the energy contained in the jet is about $10-15\%$ of the kinetic energy of the  entire ejecta to reproduce the detected profile. This study supports that the remnant originated from a jet-driven core-collapse supernova.
}

   \authorrunning{H. Yu \& J. Fang }            
   \titlerunning{ An explanation of the periphery of G309.2-0.6 }  

   \maketitle

%
%
\section{Introduction}           
\label{sect:intro}
G309.2-0.6 was identified as a SNR from the nonthermal property of its radio emission (Green~\cite{G74}; Whiteoak \& Green~\cite{WG96}). However, its age and distance are still uncertain.
A detailed study using the Australia Telescope Compact Array (ATCA) showed that it has a distance of $5.4-14.1$~kpc and an age of $(1-20)\times 10^3$~yr based on HI absorption measurements (Gaensler et al.~\cite{Gea98}).
Alternatively, based on the analysis of metal-rich, nonsolar abundance material indicated in  X-ray observations with the Advanced Satellite for Cosmology and Astrophysics {\it ASCA}, G309.2-0.6 is a young ejecta-dominated SNR with an age of $700-4000$~yr and a distance of $4\pm2$~kpc~(Rakowski et al.~\cite{Rea01}).

The radio morphology of G309.2-0.6 obtained from the ATCA observation shows a distorted shell with ears and breaks (Gaensler et al.~\cite{Gea98}). The ears to the northeast and southwest are roughly symmetric in terms of brightness and shape. Gaensler et al.~(\cite{Gea98}) argued that the ears should be produced as collimated outflows or jets from a central source interacting with the ejecta in the northeast and southwest, whereas the shell to the southeast and northwest was undisturbed.

There are evidences that SNRs can originate from jet-driven bipolar supernovae. For example, based on the detailed spatially resolved spectroscopic analyses with {\it Chandra}, Lopez et al.~(\cite{Lea13}) indicated the mean metal abundances for SNR W49B were consistent with those predicted by the models of bipolar/jet-driven core-collapse supernovae. Moreover, Lopez et al.~(\cite{Lea14}) argued that the origin of SNR 0104-72.3 was a jet-driven supernova according to the ejecta abundances derived from {\it Chandra} data.

SNR G309.2-0.6 was identified as a core collapse SNR based on its peculiar morphology and location in the Galaxy (Gaensler et al.~\cite{Gea98}; Grichener \& Soker~\cite{GS17}). For core-collapse supernovae, a large amount of gravitational energy from the central dense object can be transferred to the exploding star, and two processes, i.e., the delayed neutrino mechanism (M\"{u}ller~\cite{M16}) and jet-feedback mechanism (Soker~\cite{S16}; Bear, Grichener \& Soker ~\cite{BGS17}), have been proposed to explain the transfer channel. Recent studies indicated that the properties of core-collapse SNRs with two opposite ears on the main shell were consistent with the expectation from the jet-feedback mechanism (Bear \& Soker ~\cite{BS17,BS18}; Bear, Grichener \& Soker ~\cite{BGS17}; Grichener \& Soker~\cite{GS17}).  Assuming the jets were lunched during or shortly after the explosion, the extra kinetic energies of the ears associated with core-collapse supernovae were estimated by Grichener \& Soker~(\cite{GS17}) to be 5-15 percent of the explosion energies based on a simple geometrical assumption. Moreover, the jets lunched by the binary inside of W50 can also produce the ears observed in the remnant (Broderick et al.~\cite{Bea18}).

Three-dimensional (3D)
hydrodynamical/magnetohydrodynamical (HD/MHD) simulations are widely adopted to study the morphologies of SNRs (Orlando et al.~\cite{Oea12}; Toledo-Roy et al.~\cite{Tea14}; Fang et al.~\cite{Fea17,Fea18}). More recently, Akashi et al.~(\cite{Aea18}) indicated that the features of barrel-like and H-like shapes which are observed in planetary nebulae and SNRs can  originate from the interaction of jets and a surrounding shell. In this paper, we also investigate the dynamical evolution of G309.2-0.6 using 3D HD simulation with the assumption that the supernova ejecta has a jet component.  Two opposite ears protruding from the main shell can be produced, and the kinetic energy in the jet component can be constrained by comparison with the detected radio morphology. In Section \ref{sect:model}, the model and numerical setup are presented, and the results from simulations are given in Section \ref{sect:result}. Finally, the main conclusions and some discussion are provided in Section \ref{sect:discon}.

\section{The model and numerical setup}
\label{sect:model}

   \begin{figure}
   \centering
   \includegraphics[width=0.8\textwidth, angle=0]{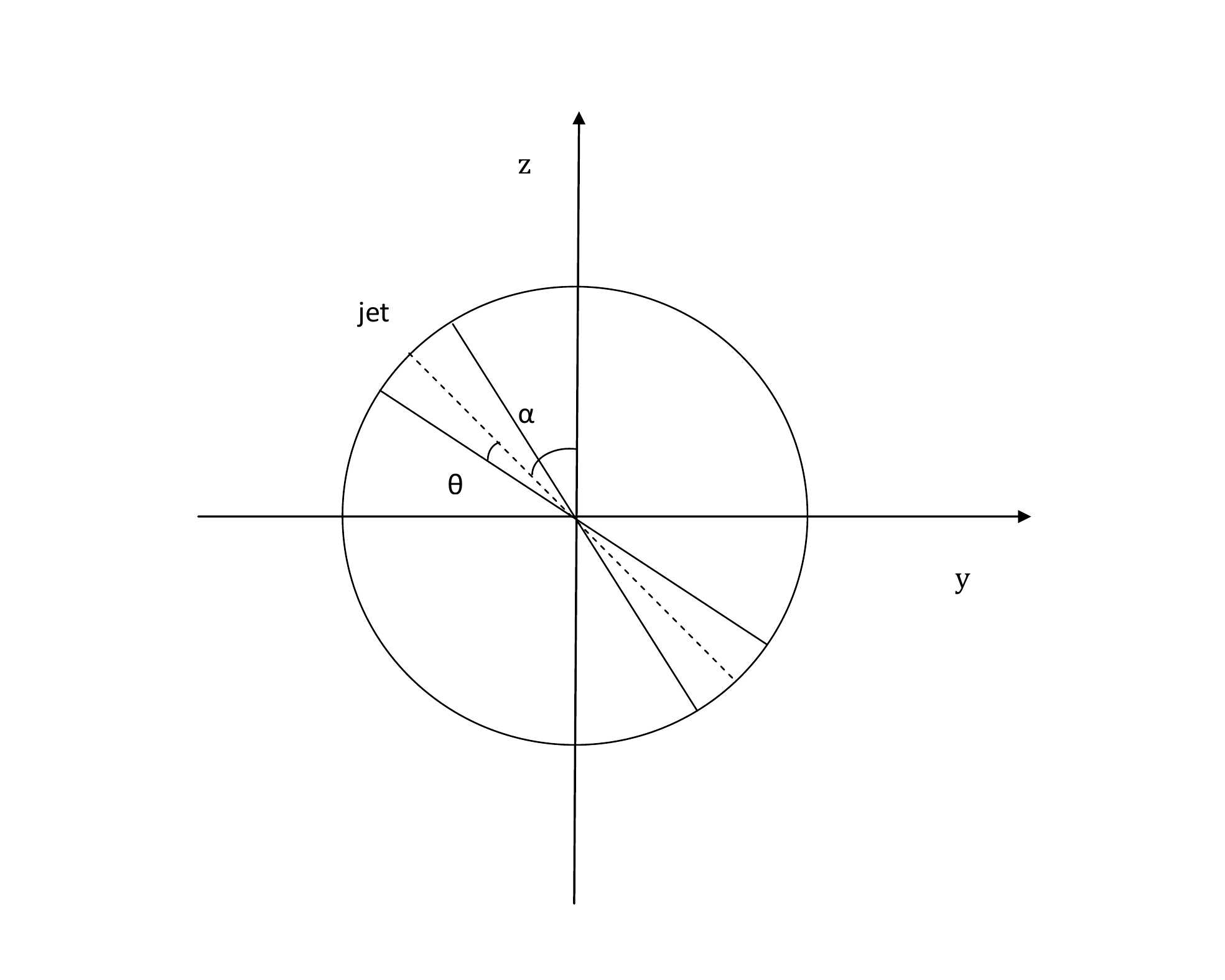}
   \caption{Scheme of the model for the supernova ejecta, which is symmetric with respect to the plane $x=0$, in the simulations. The jet component of the ejecta is inclined with an angle of $\alpha$ with respect to the $+z$ direction, and it is conical with a half apex angle of $\theta$. }
   \label{fig:jet}
   \end{figure}

The dynamical evolution of the remnant is initiated by setting supernova ejecta in an ionized medium with a density of $\rho=\mu m_{\mathrm H} n_{\mathrm H}$, where  $m_{\mathrm H}$ is the hydrogen mass, $\mu=1.4$ is the mean atomic mass  assuming  a $10:1$ H:He ratio and $n_{\mathrm H}$is the hydrogen number density. The ejecta has a mass of $M_{\mathrm{ej}}=3\,M_{\odot}$, a kinetic energy of $E_{\rm ej}=10^{51}$\,erg and a radius of $R_{\mathrm{ej}}=0.5$\,pc. The initial condition for the ejecta associated with an SNR consists of an inner core with a constant density and an outer layer with a power-law density profile. This approach is
widely adopted  to study the evolution of the remnant (Colgate \& McKee~\cite{CM69}; Jun \& Norman~\cite{JN96}; Truelove \& McKee~\cite{TM99}). The inner part of the ejecta with radius $r<r_{\mathrm c}$ is uniform, whereas the density in the outer part follows a power law on $r$ with an index of $s=9$ for the core collapse SNR (Jun \& Norman~\cite{JN96}; Truelove \& McKee~\cite{TM99}),  i.e.,
\begin{equation}
\rho_{\rm ej}(r) = \left\{
  \begin{array}{cc}
    \rho_{\mathrm c}& \mathrm{if~} r < r_{\mathrm c}\;, \\
    \rho_{\mathrm c}(r/r_{\mathrm c})^{-s} & \mathrm{if~} r_{\mathrm c} < r <  R_{\mathrm{ej}}\;,
    \label{rho_c}
   \end{array}
\right.
\end{equation}
and
\begin{equation}
r_{\mathrm c}=\left[ \frac{3-4\eta}{3(1-\eta)} \right]^{\frac{1}{s-3}} R_{\mathrm{ej}} ,
\end{equation}
where $\eta$ is the mass ratio of the outer part to that of the entire ejecta, and we adopt $\eta=3/7$ in this paper. Furthermore, the ejecta is assumed to have a jet component, which has a conical shape with an inclination angle of $\alpha$ with respect to the $+z$ direction and a half apex angle of $\theta$ (see Fig.\ref{fig:jet}), and it contains a kinetic energy of $\eta_{\mathrm{jet}} E_{\rm ej}$. Initially, both the ejecta and circumstellar matter have the same density distribution with a temperature of $T_{0}=10^4$~K, and the
velocity of the matter in the ejecta at $\mathbf{r}$ is
\begin{equation}
{\bf v} = \frac{\bf r}{R_{\rm ej}}v_{0}\;,
\end{equation}
where $v_{0}$ is the velocity of matter at the border of the ejecta. This velocity for the jet component and the other part can be calculated with
\begin{equation}
v_{0, \mathrm {jet}} = (\eta_{\mathrm{jet}}E_{\rm ej})^{1/2}\left \{\frac{2\pi \rho_{\rm c}r_{\rm c}^5}{5R_{\rm ej}^2}
+ \frac{2\pi \rho_{R}R_{\rm ej}^3\left[1-(R_{\rm ej}/r_{\rm c})^{s-5}\right]}{5-s}\right \}^{-1/2}\;,
\end{equation}
and
\begin{equation}
v_{0, \mathrm {ms}} = [(1.0 - \eta_{\mathrm{jet}})E_{\rm ej}]^{1/2}\left \{\frac{2\pi \rho_{\rm c}r_{\rm c}^5}{5R_{\rm ej}^2}
+ \frac{2\pi \rho_{R}R_{\rm ej}^3\left[1-(R_{\rm ej}/r_{\rm c})^{s-5}\right]}{5-s}\right \}^{-1/2}\;,
\end{equation}
respectively, where $\rho_{R}$ is the density of ejecta at $R_{\rm ej}$.
$t=0$ corresponds to the age of ejecta with a radius $R_{\rm ej}$ after the supernova.
The ratio of the mass contained in the jet component to that of the entire ejecta is $\eta_{\rm m}= 1-\cos\theta$.
The other details of the velocity and the density of the materials in the ejecta can be seen in Jun \& Norman~(\cite{JN96}).

Neglecting the radiative cooling and particle acceleration involved in the SNR, its dynamical evolution can be derived based on the Euler equations, i.e.,
\begin{eqnarray}
\frac{\partial\rho}{\partial t} + \nabla\cdot(\rho \textbf{v}) & = & 0\; , \\
\frac{\partial \rho {\bf v}}{\partial t} + \nabla \cdot ( \rho {\bf
    vv} )   + \nabla{P} & = & 0\; , \\
\frac{\partial E}{\partial t} + \nabla \cdot (E+P){\bf v} ) & = & 0 ,
\end{eqnarray}
where $P $ is the gas pressure, $E$ is the total energy density
\begin{equation}
E = \frac{P}{\gamma - 1}+\frac{1}{2}\rho v^2 \;,
\nonumber
\end{equation}
and $t$ is time. The adiabatic index $\gamma$ is adopted to be $5/3$
for the nonrelativistic gas, and $\bf{v}$ is the gas velocity. These equations are solved based on the PLUTO code (Mignone et al.~\cite{Mea07,Mea12}) in a 3D Cartesian coordinate system. The simulations are performed in a cubic domain of $24\times24\times24$~pc$^{3}$ with an equivalent $512\times512\times512$ grid cells.

\section{Results}
\label{sect:result}

\begin{table}
\begin{center}
\caption[]{ Parameters for the Different Models in the Simulations. The Common Parameters are $E_{\rm ej}=10^{51}\,\mathrm{erg}$, $M_{\mathrm{ej}}=3\,M_{\odot}$, $R_{\mathrm{ej}}=0.5$\,pc, $n=1\mathrm{cm}^{-3}$, $\eta=3/7$, $T_{0}=10^4$~K and $\alpha=50^{\circ}$.  }\label{Tab:para}
 \begin{tabular}{ccccccc}
  \hline\noalign{\smallskip}
Parameters             & Model A  &   Model B  & Model C & Model D &   Model E  & Model F           \\
  \hline\noalign{\smallskip}
$\theta(^{\circ})$     & 10       &     10     &   10    &     10   &   10    &     15  \\
$\eta_{\mathrm{jet}}$  & 0.05     &     0.1    &   0.15  &     0.3   &   0.4    &     0.3 \\
  \noalign{\smallskip}\hline
\end{tabular}
\end{center}
\end{table}

\subsection{Dynamical Evolution of the Ejecta with a Jet Component}

\begin{figure}
   \centering
   \includegraphics[width=1.0\textwidth, angle=0]{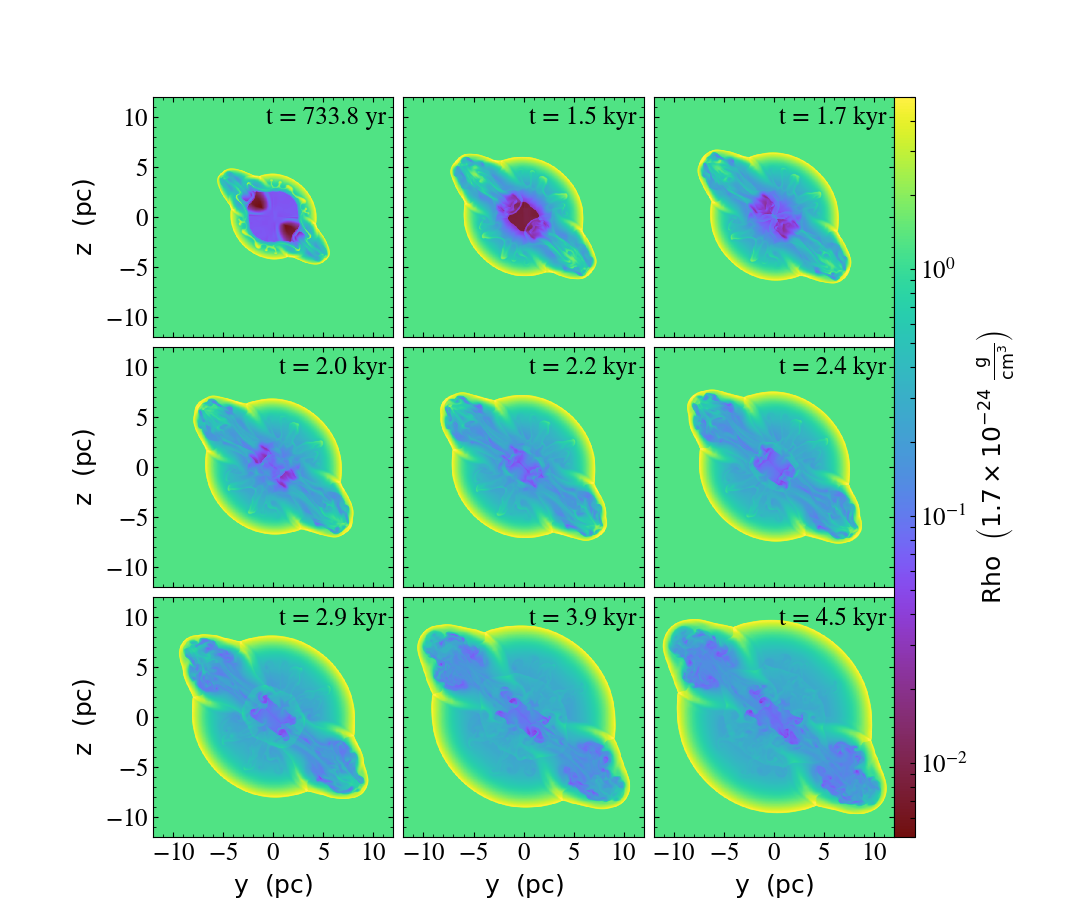}
   \caption{Slices of the density in the plane $x=0$ at different evolution times for Model C with the parameters $E_{\rm ej}=10^{51}\,\mathrm{erg}$, $M_{\mathrm{ej}}=3\,M_{\odot}$, $R_{\mathrm{ej}}=0.5$\,pc, $\eta=3/7$, $T_{0}=10^4$~K, $\alpha=50^{\circ}$, $\theta=10^{\circ}$ and $\eta_{\mathrm{jet}}=0.15$. }
   \label{fig:rho}
   \end{figure}
\begin{figure}
   \centering
   \includegraphics[width=1.0\textwidth, angle=0]{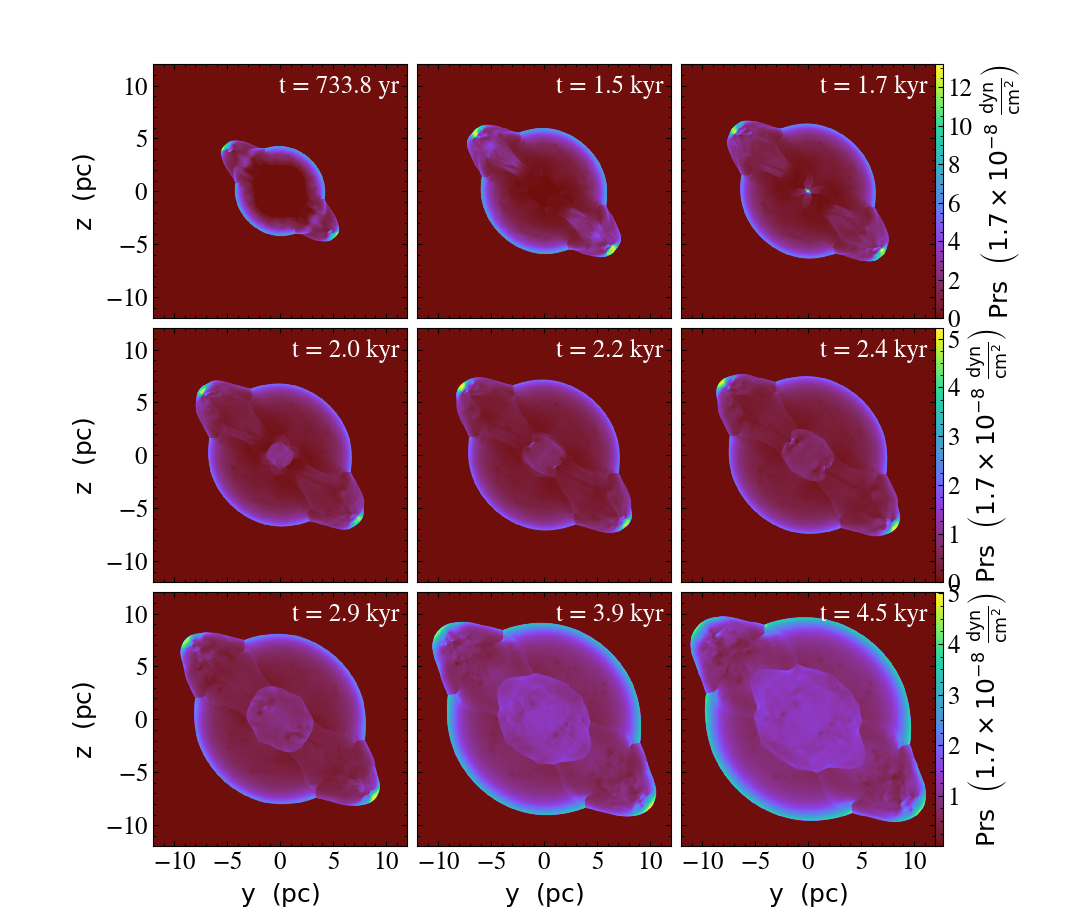}
   \caption{Slices of the pressure in the plane $x=0$ at different evolution times for Model C. }
   \label{fig:pre}
   \end{figure}

   \begin{figure}
   \centering
   \includegraphics[width=1.0\textwidth, angle=0]{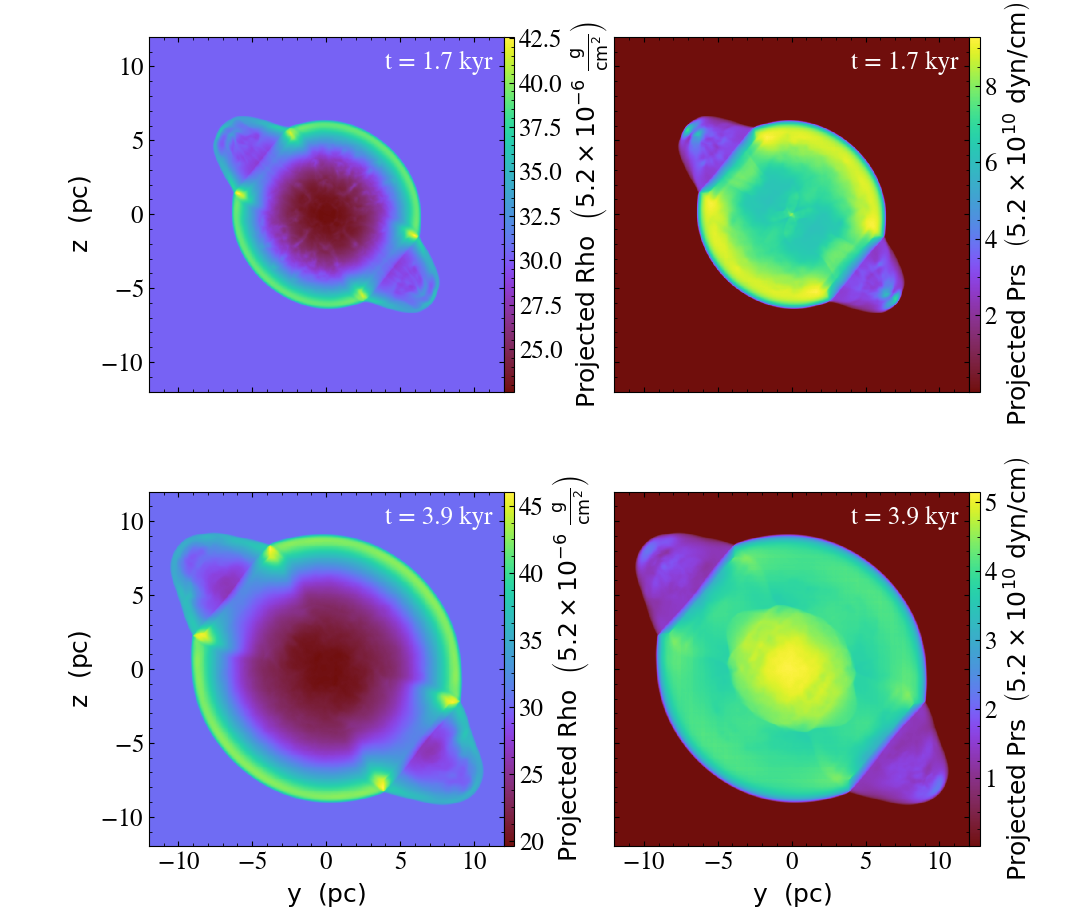}
   \caption{The projected density (left) and pressure (right) along $x$, i.e., the integral of the field with respect to $x$ from $-12$~pc to $12$~pc, at the evolution times $1.7\times10^3$~yr (top) and $3.9\times10^3$~yr (bottom) for Model C. }
   \label{fig:rhopj}
   \end{figure}

Fig.\ref{fig:rho} and Fig.\ref{fig:pre} show the distributions of  density and pressure, respectively, in the plane $x=0$ from the simulation with parameters for Model C in Table \ref{Tab:para}. 

In Model C with $\theta=10^{\circ}$, the density for the inner part of the entire ejecta is $\rho_{\mathrm{c}}=1.54\times10^2 n_{\mathrm{H}}{\mathrm{cm}}^{-3}$ with $r_{\mathrm{ c}}= 0.48\mathrm{pc}$, and the mass in the jet is $0.015 M_{\mathrm{ej}}$ . The simulation begins with $r_{\rm ej}=0.5\mathrm{pc}$, which corresponds to an age of $r_{\rm ej}/v_{0, \mathrm{ms}} = 56$~yr (see Eq. (9) in Truelove \& McKee~(\cite{TM99})) for the remnant.  Initially, $v_0$, i.e., the velocity of matter at the border of eject, is $2.9\times10^9\mathrm{cm\, s}^{-1}$ in the jet and $8.7\times10^8\mathrm{cm\, s}^{-1}$ outside the jet. This matter expands into the ambient uniform medium in which the sound speed is $1.5\times10^6 \mathrm{cm\, s}^{-1}$, and then a forward shock ahead of the ejecta is generated due to the supersonic motion.  The ambient matter is compressed and thermalized by the forward shock. As a result, the thermalized medium drives a reverse shock which continually compresses the matter in the ejecta. As the forward shock expands outwardly, the material in the inner part of the remnant becomes more and more tenuous  until it encounters  the reverse shock. Rayleigh-Taylor instabilities develop near the location of the contact discontinuity which is the border between the shocked circumstellar medium and the shocked ejecta.

In the jet direction, the reverse shock propagates more quickly towards the center of the simulation as compared with the main shell outside the two bumps. At a time of $\geq2.0\times10^3$~yr, the northeast and southwest components of the reverse shock are encountered at the simulation center. 

In Fig.\ref{fig:pre}, the reverse shock is reflected after the encounter; the shocked ejecta is further thermalized by the reflected reverse shock, and an oval shape is clearly indicated in a snapshot of the pressure in the plane $x=0$ after $2.0\times10^3$~yr.

The projected density along $x$, i.e., the integral of density with respect to $x$ from $-12$~pc to $12$~pc., for $t=1.7\times10^3$~yr (top) and $t=3.9\times10^3$~yr (bottom)  is illustrated in Fig.\ref{fig:rhopj}. Most of the material inside the remnant is located at the main shell, and the projected density at the border of the jet is higher than in other parts of the shell. The radio emission from G309.2-0.6 is concentrated in the distorted shell. Moreover, the intersection of the northeast bump and the main shell, and the shell to the southwest of the jet are more luminous than other parts of it (Gaensler et al.~\cite{Gea98}). Although the derivation of radio morphology from the simulation needs more information on the distribution of relativistic electrons and magnetic field in the remnant, the shell structure with jets as indicated in radio is reproduced in the morphology of the projected density.

In the right panels of Fig.\ref{fig:rhopj}, we also show the projected pressure of the remnant along the $x$ direction. Before the collision of the reverse shock at the center, the pressure around the center is insignificant compared with that near the shell. After the collision, the ejecta inside the remnant is re-shocked by the reflected reverse shock, and the pressure in the oval region around the center of the remnant will become more significant than in the shell.

 \begin{figure}
   \centering
   \includegraphics[width=0.7\textwidth, angle=0]{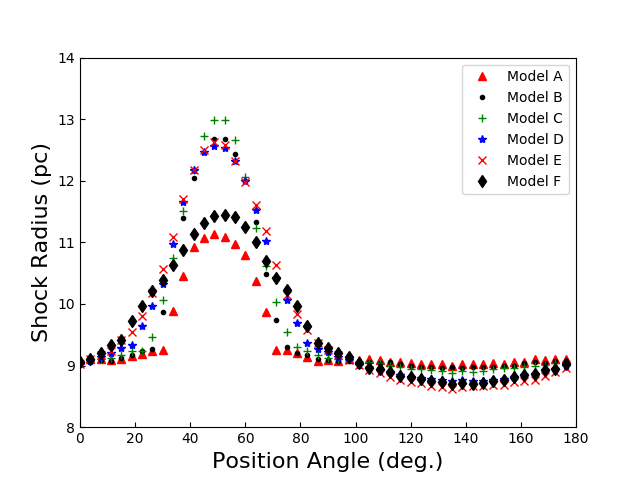}
   \caption{Radius of the forward shock at $t=3.9\times10^3$~yr in relation to the position angle, i.e., the angle east of north with respect to the simulation center, in the plane $x = 0$ for the different models. }
   \label{fig:radius}
   \end{figure}

Due to the anisotropy of the supernova ejecta with respect to a jet component, the radius of the forward shock varies with direction. At the position angles, i.e., the angle east of north with respect to the center of the simulation, $\phi\sim50^{\circ}$ and $230^{\circ}$ for an age of $t=3.9\times10^3$~yr, the maximum radius is  $\sim13$~pc. For the  shell, excluding the bumps to the northeast and southwest, the radius, which is around $\sim9$~pc, varies slowly with position angle. Therefore, the ratio $\eta_{\mathrm{jm}}$ of the radius in the jet direction to that for the main shell excluding the two bumps is $\sim1.4$.

\subsection{Periphery of the SNR with Different Jet Configuration}
   \begin{figure}
   \centering
   \includegraphics[width=1.0\textwidth, angle=0]{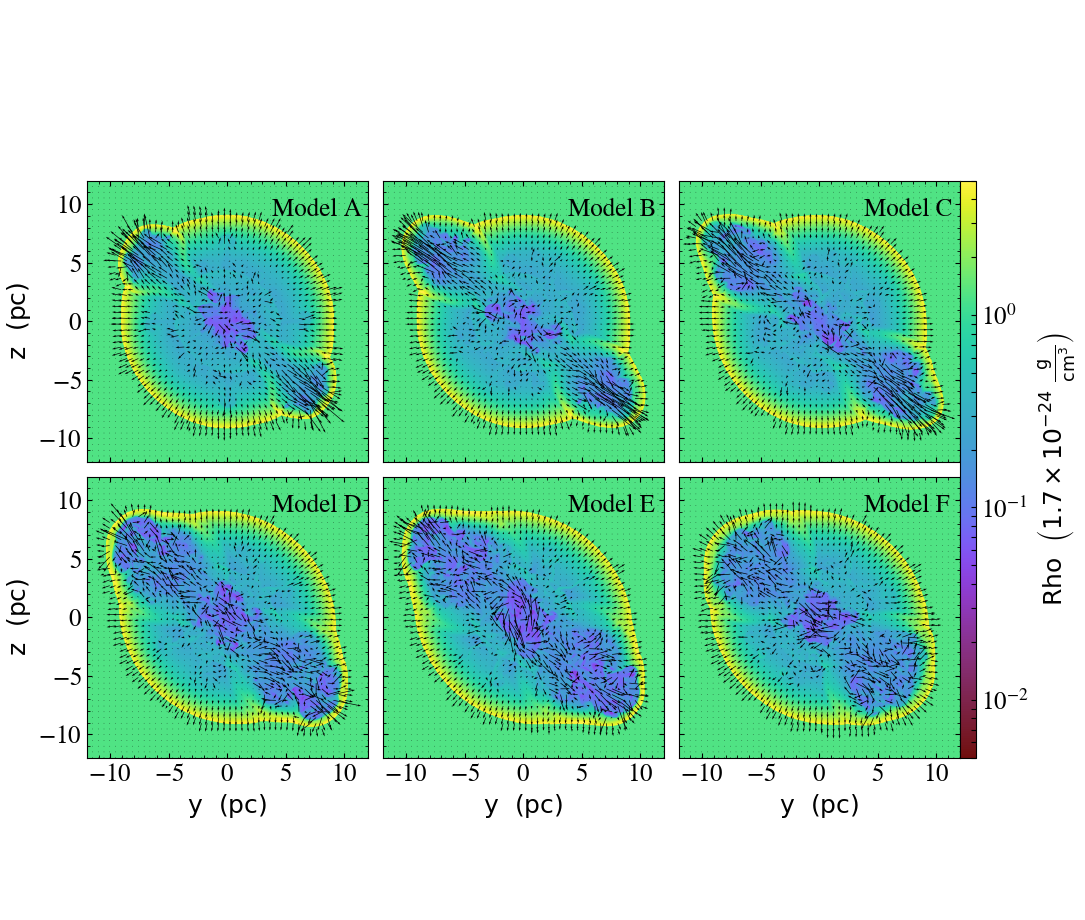}
   \caption{Slices of the density with the velocity vectors in the plane $x=0$ at $t=3.9\times10^3$~yr for the models from A to F, respectively.}
   \label{fig:abcd}
   \end{figure}

With $\theta=10^{\circ}$ and $\eta_{\mathrm{jet}}=0.15$, the ratio of the mass contained in the jet to the entire ejecta is $1.52\%$, and the velocity of the forward shock in the jet is larger than the main shell by a factor of $1.5$. Consequently, two opposite prominent bumps in the jet direction protrude on the SNR shell. As indicated in Fig.\ref{fig:radius}, at the simulation time of $3900\,\mathrm{yr}$, the radiuses of the forward shock in the jet direction and the shell excluding the two bumps, are $\sim13$~pc and $\sim9$~pc, respectively, which are consistent with the detected values as indicated in the radio observation for SNR G309.2-0.6 with a distance of $\sim 5$~kpc.

Fig.\ref{fig:abcd} shows the resulting snapshot of  the density in the plane $x=0$ for the different models.  With a smaller $\eta_{\mathrm{jet}}$ of 0.05 (Model A), the maximum distance of the border of the forward shock in the jet direction to the center of the simulation is 11.1~$\mathrm{pc}$, and that with the position angle between $100 - 160$ is also $\sim 9$~pc. As indicated in Fig.\ref{fig:radius}, the extension of the position angle of the jet in the snapshot of the plane $x=0$ is also related with $\eta_{\mathrm{jet}}$. In Model C, the angular extension of the bump with the distance of the border to the center of the simulation larger than $9.5$~pc is $\sim 45^{\circ}$, whereas it becomes $\sim 56^{\circ}$ in Model E with $\eta_{\mathrm{jet}}=0.4$. In Model F with $\eta_{\mathrm{jet}}=0.3$ and $\theta=15^{\circ}$, the jet component of the ejecta which is more extended in the position angle has a lower initial velocity compared with Model D with a ratio of $(1-\cos(10^{\circ})/(1-\cos(15^{\circ})))^{1/2}=0.67$. As a result, at $t=3.9\times10^3$~yr, the bumps are inconspicuous with regard to distances to the center of the simulation not much larger than the main shell in Model F.

%

\section{Discussion and conclusions}
\label{sect:discon}

In this paper, we investigate the reason for the formation of the peculiar periphery as indicated in the radio observations for SNR G309.2-0.6 based on 3D HD simulation. In the model, soon after the supernova, the ejecta contains a jet component which has a higher velocity than the other part. During the
evolution of the ejecta in the uniform circumstellar medium, two bumps are formed on the main shell. The resulting profile of the forward shock can be used to constrain
the age, energy ratio $\eta_{\mathrm{jet}}$ and  half opening angle of the jet component $\theta$ if the mass, kinetic energy of the ejecta, density
of the ambient medium and distance to the remnant are known.

Assuming SNR G309.2-0.6 has a distance of $5$~kpc, the radius of the
main shell outside the two bumps is $\sim 9$~pc, which is consistent with Model A to C at the simulation time of $\sim3900$~yr with $E_{\rm ej}=10^{51}\,\mathrm{erg}$, $M_{\mathrm{ej}}=3\,M_{\odot}$ and $n=1\mathrm{cm}^{-3}$. This age is in the range of $(1-20)\times 10^3$~yr derived from the HI absorption measurements, and it is also consistent with the age of $\leq 4000$~yr under the assumption that the proposed outflow from the remnant is interacting with the HII region RCW 80 (Gaensler et al.~(\cite{Gea98}).
With $\theta=10^{\circ}$, we find the resulting profile of the remnant in Models B or C is similar to that from the radio image.  Therefore, an energy ratio of about $0.1 - 0.15$ is appropriate for SNR G309.2-0.6 to explain its peculiar periphery, and our results support the assumption of the jet model.

Recently, Grichener \& Soker (\cite{GS17}) investigated the kinetic energy of the jets which induce ears on core collapse SNRs, and it is about $5-15$ percent of the explosion energy for those remnants with ears based on  simple geometrical assumptions which ignore the details of evolution. Especially, for  SNR G309.2-0.6, the ratio of the kinetic energy to the entire shell was estimated to be $\sim 7\%$ (Grichener \& Soker~\cite{GS17}). In this paper, using 3D HD simulation, the ratio is roughly constrained to be $\sim 10 - 15$ percent with $\theta=10^{\circ}$, which is consistent with that derived in Grichener \& Soker~(\cite{GS17}) for core collapse SNRs.

\begin{acknowledgements}
 HY is partially supported by the Yunnan Applied
Basic Research Project (2016FD105), ang the Foundations of Yunnan Province (2016ZZX180 and 2016DG006) and Kunming University (YJL15004 and XJL15015).
JF is partially supported by the National
Natural Science Foundation of China (NSFC) under No.11563009,
the Yunnan Applied Basic Research
Project (2016FB001), the Candidate Talents Training Fund of Yunnan Province (2017HB003)
and the Program for Excellent Young Talents, Yunnan University (WX069051 and 2017YDYQ01).
\end{acknowledgements}

\label{lastpage}

\end{document}